\documentclass[a4paper]{article}

\usepackage{INTERSPEECH2022}
\usepackage{multirow}
\usepackage{diagbox}

\title{Hierarchical Attention Network for Evaluating Therapist Empathy in Counseling Session}
\name{Dehua Tao$^1$, Tan Lee$^1$, Harold Chui$^2$, Sarah Luk$^2$}
\address{
  $^1$ Department of Electronic Engineering \quad 
  $^2$ Department of Educational Psychology\\The Chinese University of Hong Kong}
\email{dhtao@link.cuhk.edu.hk, tanlee@ee.cuhk.edu.hk, \{haroldchui, sarah\_luk\}@cuhk.edu.hk}

\begin{document}

\maketitle
\begin{abstract}
Counseling typically takes the form of spoken conversation between a therapist and a client. The empathy level expressed by the therapist is considered to be an essential quality factor of counseling outcome. This paper proposes a hierarchical recurrent network combined with two-level attention mechanisms to determine the therapist's empathy level solely from the acoustic features of conversational speech in a counseling session. The experimental results show that the proposed model can achieve an accuracy of $72.1\%$ in classifying the therapist's empathy level as being ``high" or ``low". It is found that the speech from both the therapist and the client are contributing to predicting the empathy level that is subjectively rated by an expert observer. By analyzing speaker turns assigned with high attention weights, it is observed that $2$ to $6$ consecutive turns should be considered together to provide useful clues for detecting empathy, and the observer tends to take the whole session into consideration when rating the therapist empathy, instead of relying on a few specific speaker turns.
\end{abstract}

\section{Introduction}

Psychotherapy is typically conducted through conversation between a therapist and a client, aiming to help the client change behavior, overcome difficulties and/or manage distress. In the field of psychotherapy, empathy is described as ``the therapist's sensitive ability and willingness to understand the client's thoughts, feelings, and struggles from the client's point of view" \cite{rogers1995way}. Previous studies have shown that therapist empathy is positively associated with client outcomes and is considered an important quality indicator of psychotherapy \cite{miller2009toward,elliott2011empathy,moyers2013low,elliott2018therapist}. Empathy level of the therapist is often evaluated in a subjective manner, either by the therapist (empathic resonance with the client), the observer (expressed empathy), or the client (received empathy) \cite{elliott2011empathy}. However, subjective assessment is time-consuming and costly. The present study aims to investigate methods of automatically analyzing conversational speech in counseling sessions and determining the therapist's empathy level. The outcome of objective evaluation would be useful to the training and improvement of counseling skills.

Spoken language carries a wide range of information, in which one of the major dimensions is determined by and related to the speaker's emotion, desire, intent, and mental state. It was shown that the empathy level expressed by the therapist could be predicted from language cues found in the transcripts of therapy-client interactions \cite{xiao2012analyzing,chakravarthula2015assessing, gibson2015predicting,gibson2016deep}. As the primary medium of communication between the therapist and the client, the acoustic speech signal is believed to contain pertinent information about the speaker state. For example, the vocal cues, including pitch, intensity, duration, and speech rate, were investigated in the analysis of therapist empathy \cite{xiao2013modeling,xiao2014modeling,imel2014association,xiao2015analyzing}. These investigations were focused on the relation between low-level speech features and therapist empathy. In this paper, we present a deep neural network (DNN) based system to determine the therapist's empathy level from acoustic properties of speech in the therapist-client conversation.

In a typical counseling session, a therapist and a client take turn to speak, constituting a long conversation. A speaker turn refers to the time period during which only one person speaks. A counseling session may contain hundreds of speaker turns. A speaker turn can be further divided into sub-turns for analysis and feature extraction. In this way, each counseling session is represented by a hierarchical structure, i.e., a group of sub-turns constitute a turn, many speaker turns make up the session. A hierarchical recurrent neural network is adopted to model speech features at different levels of the hierarchy. It is noted that different sub-turns and turns carry different importance weights in determining the therapist empathy. Therefore two levels of attention mechanisms \cite{bahdanau2014neural,luong2015effective}, one at the sub-turn level and the other at the turn level, are incorporated to improve the performance of the model. To our best knowledge, this is the first use of a hierarchical recurrent neural network combined with two-level attention mechanisms (HRAN) for evaluating the therapist empathy from conversational speech between the therapist and the client.

Experimental results show that the proposed system is able to determine whether the therapist's empathy level is high or low solely from the speech signals in a given counseling session. It is found that the speech from both the therapist and the client contribute to predicting the third-party observer's subjective rating. By analyzing the speaker turns assigned with high attention weights, it is suggested that consecutive turns should be considered in conjunction, and the observer takes a holistic approach in empathy rating instead of relying on specific parts of the conversation.

Section \ref{sec:data} describes a large-scale counseling speech database used in this research. Section \ref{sec:HRAN} introduces the proposed HRAN model for empathy classification in detail. Section \ref{sec:exp} presents the experiments and results, followed by discussion and conclusion in Section \ref{sec:disc} and Section \ref{sec:conc}, respectively.

\section{Speech database of counseling}
\label{sec:data}

\begin{table*}[htb]
  \caption{Summary of the counseling speech data used in this study}
  \label{tab:data}
  \centering
  \resizebox{1.0\textwidth}{!}{
  \begin{tabular}{|c|c|c|c|c|c|c|}
  \hline
          & \begin{tabular}[c]{@{}c@{}}Average speech time \\ per session (min)\end{tabular} & \begin{tabular}[c]{@{}c@{}}Average no. of speaker turns \\ per session\end{tabular} & \begin{tabular}[c]{@{}c@{}}Average no. of sub-turns \\ per speaker turn\end{tabular} & \begin{tabular}[c]{@{}c@{}}Average duration \\ of turns (sec)\end{tabular} & \begin{tabular}[c]{@{}c@{}}Average duration \\ of sub-turns (sec)\end{tabular} & \begin{tabular}[c]{@{}c@{}}Average TES score \\ of sessions\end{tabular} \\ \hline
Therapist & 14.89                                                                            & 139                                                                                 & 2                                                                                    & 6.03                                                                       & 3.25                                                                           & \multirow{2}{*}{38.64}                                                   \\ \cline{1-6}
Client    & 33.66                                                                            & 138                                                                                 & 3                                                                                    & 12.93                                                                      & 3.59                                                                           &                                                                          \\ \hline
  \end{tabular}}
\end{table*}

The speech data used in this study are from audio recordings made in the counseling practicum for therapist trainees at the Department of Education Psychology, the Chinese University of Hong Kong (CUHK). Clients of the counseling practicum were adults who sought counseling assistance over various concerns related to emotion, stress, relationship, career, personal growth, and self-esteem. All therapists and clients spoke Hong Kong Cantonese. The study was approved by the institutional review board, and informed consent was obtained from both the clients and therapists.

% Before the counseling sessions, the trainees discussed the goals and strategies with their supervisors.

Each counseling session was about 50 minutes long. The therapist's empathy level in each of the counseling sessions was rated by trained raters according to the Therapist Empathy Scale (TES) \cite{decker2014development}. The TES is a nine-item observer rating scale to assess affective, cognitive, attitudinal, and attunement aspects of therapist empathy. Each item is rated on a seven-point scale from 1 $=$ \textit{not at all} to 7 $=$ \textit{extremely}. The total score (range from 9 to 63) is used in this research, with a higher score indicating higher therapist empathy. As a reliability check, about 40\% (61 sessions) of the collected sessions were rated by two raters. The intraclass coefficient was 0.90, indicating excellent interrater reliability \cite{cicchetti1994guidelines}.

% As a reliability check, about 40\% (61 sessions) of the videotapes were rated by two raters. The ICC based on a mean-rating (k = 2), consistency, two-way random effects model was .90, indicating excellent interrater reliability beyond the training phase.      

% TES scores in our data range from 9 to 63, with larger values indicating higher therapist empathy. 

% TES comprises nine component scores, each being given on a seven-point scale. The total of the nine component scores is used in this research to represent the ground-truth empathy level of the therapist. The ground-truth scores in our data range from 9 to 63, with larger values indicating higher therapist empathy. 

% During the rating process, the rater observed the therapist-client interactions while paying more attention to the therapist's speech and behavior than to the client.

The audio recording of a session was divided into speaker-turn-based audio segments according to the transcription and speaker turn markers obtained manually. The turn-level speech-text alignment is done automatically as described in \cite{Tao2022CharacterizingTS}. Each speaker turn is then divided into sub-turns, which are separated by major pauses of 0.5 second or longer. In this work, a total of 118 counseling session recordings from 39 different pairs of therapists and clients are used, with TES scores on the two extremes. According to their TES scores, the sessions are divided into two groups. The high-empathy group consists of 61 sessions and the low-empathy group has 57 sessions. TES scores of high-empathy sessions range from 42 to 56.5 (46.34\textpm3.58), and those of low-empathy sessions range from 18 to 36 (30.40\textpm4.79). Table~\ref{tab:data} gives a summary of the speech data used in this study.

\section{Proposed system}
\label{sec:HRAN}

\begin{figure}[htb]
  \centering
  \includegraphics[width=\linewidth]{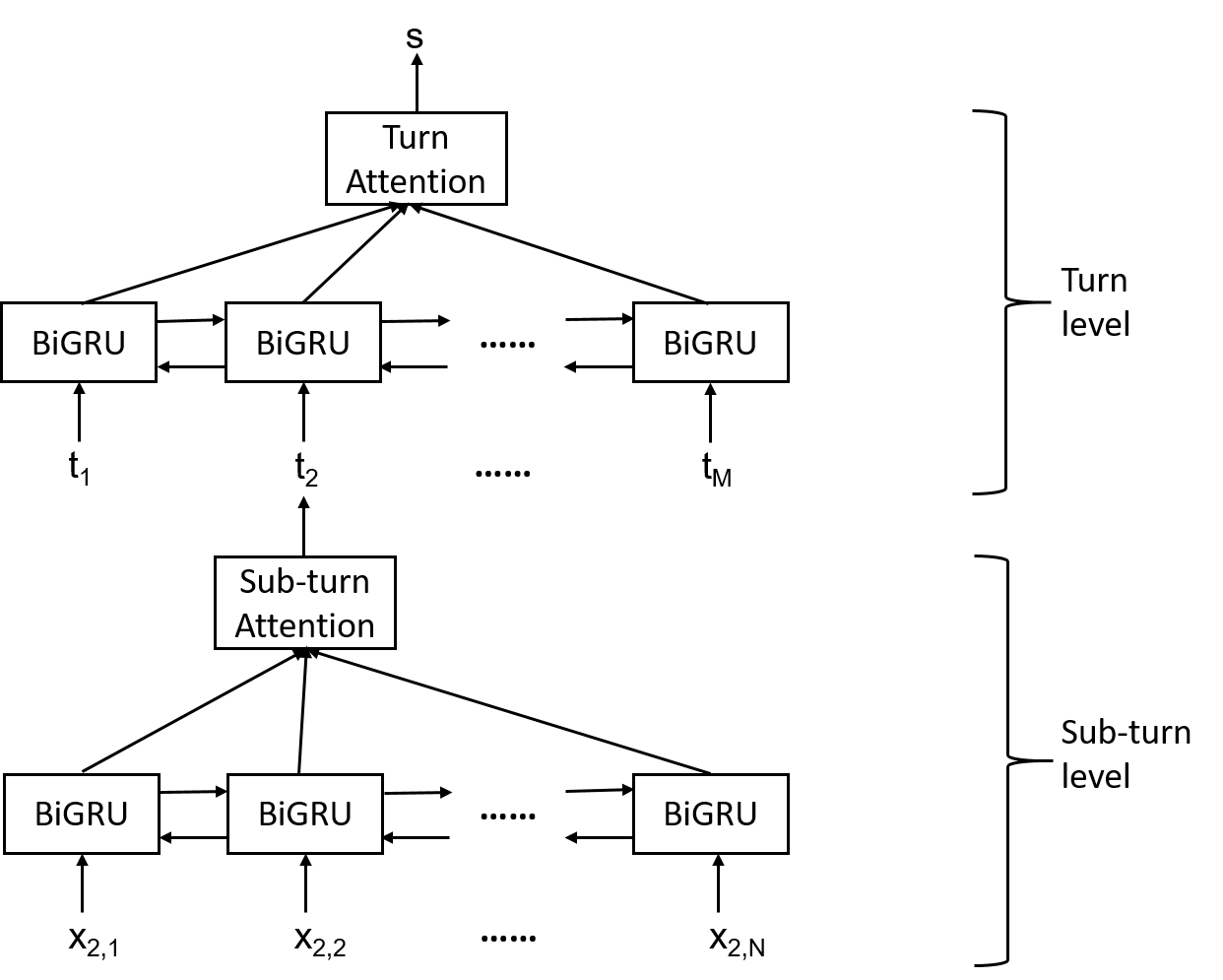}
  \caption{Architecture of the proposed system: HRAN.}
  \label{fig:HRAN}
\end{figure}

The proposed model architecture is shown as in Figure~\ref{fig:HRAN}. Consider a counseling session that comprises $M$ speaker turns, denoted by $t_i$, $i = 1, 2, ..., M$. The speaker turn $t_i$ is divided into $N_i$ sub-turns denoted as $x_{i,j}$, where $j = 1, 2, .., N_i$. In the following sub-sections, we will explain how to encode a long counseling session by a high-level embedding learned from acoustic speech features and use this embedding to classify the session as high empathy or low empathy.

\subsection{From sub-turn speech features to turn-level embedding}
\label{ssec:sub-turn}

Acoustic parameters defined by the eGeMAPS feature set \cite{eyben2015geneva} are computed from the speech signal in each sub-turn. eGeMAPS has been applied to numerous paralingusitc tasks, including emotion recognition \cite{eyben2015geneva}, speech intelligibility prediction \cite{Xue2019}, and detection of Alzheimer’s dementia \cite{haider2019assessment}. Computation of the 88-dimensional eGeMAPS feature vector is implemented by the openSMILE toolkit \cite{eyben2010opensmile} following the default configuration. 

A bidirectional gated recurrent unit (GRU) \cite{cho2014properties} is used to convert the sub-turn acoustic feature vector into an sub-turn-level embedding. The bidirectional GRU makes use of contextual information from both forward and backward time directions. The hidden state $h_{i,j}$ for sub-turn $x_{i,j}$ is obtained by concatenating the forward GRU state $\overrightarrow{h_{i,j}}$, which reads turn $t_i$ from $x_{i,1}$ to $x_{i,N}$, and the backward GRU state $\overleftarrow{h_{i,j}}$, which reads from $x_{i,N}$ to $x_{i,1}$.

\begin{equation}
\begin{aligned}
  &\overrightarrow{h_{i,j}} = \overrightarrow{\text{GRU}}(x_{i,j}, h_{i,{j-1}}) \\
  &\overleftarrow{h_{i,j}} = \overleftarrow{\text{GRU}}(x_{i,j}, h_{i,{j+1}}) \\
  &h_{i,j} = [\overrightarrow{h_{i,j}}, \overleftarrow{h_{i,j}}]
  \label{eq1}
\end{aligned}
\end{equation}

To model the non-uniform contributions from individual sub-turns, an attention mechanism is included in the model,

\begin{equation}
\begin{aligned}
  &u_{i,j} = \text{tanh}(W_xh_{i,j} + b_x) \\
  &\alpha_{i,j} = \frac{\text{exp}(u^\top_{i,j}u_x)}{\sum_j \text{exp}(u^\top_{i,j}u_x)} \\
  &t_i = \sum_j \alpha_{i,j}h_{i,j}
  \label{eq2}
\end{aligned}
\end{equation}

A one-layer Multi-layer Perceptron (MLP) is first used to generate $u_{i,j}$ as a hidden representation of $h_{i,j}$. The importance of the sub-turn is measured as the similarity between $u_{i,j}$ and a sub-turn-level vector $u_x$. A softmax function is applied to compute the normalized importance weight $\alpha_{i,j}$. Lastly, a turn-level embedding $t_i$ is obtained as a weighted sum of sub-turn GRU hidden states. The learnable vector $u_x$ has the same dimension as $u_{i,j}$ and is initialized randomly. It is used as a query ``which sub-turn is related to therapist empathy" over the sub-turns \cite{sukhbaatar2015end}.

\subsection{From turn-level embeddings to session embedding}

Given the turn-level embeddings, the session-level embedding is obtained using a similar approach. A bidirectional GRU is used again to obtain the hidden states for the speaker turns:

\begin{equation}
\begin{aligned}
  &\overrightarrow{h_i} = \overrightarrow{\text{GRU}}(t_i, h_{i-1}) \\
  &\overleftarrow{h_i} = \overleftarrow{\text{GRU}}(t_i, h_{i+1}) \\
  &h_i = [\overrightarrow{h_i}, \overleftarrow{h_i}]
  \label{eq3}
\end{aligned}
\end{equation}

The hidden state $h_i$ for turn $t_i$ is derived based on the context information from both forward and backward directions. In order to assign different importance weights related to therapist empathy, an attention mechanism is incorporated via a turn-level vector $u_t$ measuring the importance of individual turns, i.e.,

\begin{equation}
\begin{aligned}
  &u_i = \text{tanh}(W_th_i + b_t) \\
  &\alpha_i = \frac{\text{exp}(u^\top_iu_t)}{\sum_i \text{exp}(u^\top_iu_t)} \\
  &s = \sum_i \alpha_ih_i
  \label{eq4}
\end{aligned}
\end{equation}

$s$ is a session-level embedding derived from all turns in the session. Similarly, $u_t$ is a learnable vector that can be initialized randomly.

For binary classification of high-low empathy level, the session embedding $s$ is fed into a dense layer, followed by a softmax layer:

\begin{equation}
\begin{aligned}
  p = \text{softmax}(W_ss + b_s)
  \label{eq5}
\end{aligned}
\end{equation}

In our experiments, the training loss is defined as the negative log likelihood of the correct class,

\begin{equation}
\begin{aligned}
  L = -\sum_k\text{log}p_{k,c}
  \label{eq6}
\end{aligned}
\end{equation}

where $c$ is the class label of the session $k$.

\section{Experiments and Results}
\label{sec:exp}

\subsection{Experimental setup}

6-fold cross-validation (CV) experiment is arranged by splitting the 118 counseling sessions into three sub-sets, i.e., training (4 folds), development (1 fold) and test (1 fold) set. The input data for empathy level classification are the conversational speech recorded in counseling sessions. The classification output is binary, indicating the therapist empathy level as ``high" or ``low". There are two speakers in each counseling session. The sub-turn-level eGeMAPS features from each speaker are normalized with respect to the mean and variance computed from all sub-turn utterances from the speaker in the whole session. 

The hidden layer sizes of GRU used at the sub-turn and turn levels are 64 and 16, respectively. The query vector in the attention layer has the same dimension as the hidden state of bidirectional GRU. The model is trained for a fixed number of epochs (50) and evaluated on the development set after each epoch. The best model is selected based on the classification accuracy on the development set. The model parameters are updated with the speech data of one session at each iteration. An Adam optimiser with $\beta_1 = 0.9$, $\beta_2 = 0.999$ and initial learning rate of 0.0001 is used. The learning rate is reduced by 0.1 every 30 epochs. The average of the classification accuracies on the 6-fold CV is used to indicate the model performance.

In order to evaluate the effectiveness of the two-level attention mechanisms in the model, three baseline models are investigated in the experiment: a hierarchical recurrent neural network without attention (HRN), a hierarchical recurrent neural network with only the sub-turn-level attention (HRSAN) and one with only the turn-level attention (HRTAN). The three models have the same architectures and hyper-parameters as the HRAN, except for the attention layers. Model evaluation is carried out with speech data from either one, or both the therapist and the client.

% In order to prove the efficiency of the HBGAN in our task, three baseline models are designed for comparison: a bidirectional GRU network (BGN), a bidirectional GRU attention network (BGAN), and a hierarchical bidirectional GRU network (HBGN). Since turns have been split into sub-turns, the BGN treats the whole session as a sequence of the sub-turns. The input is the sub-turns in the joint set of all the turns $t_i$ with $i\in[1, M]$ in the session. Following the bidirectional GRU layer with the hidden units of 64 is a dense layer with the output size of 32, which can keep the dimension of final session acoustic embedding same as that in the HBGAN. BGAN has the same setting as the BGN, plus attention mechanism same as that in the sub-turn level of the HBGAN. The two networks are designed to test the benefit of having the hierarchical mechanism when dealing with the dialogue data. The HBGN has the same structure with the HBGAN but without the attention mechanism. This network is used to test the benefit of having the attention mechanism.

\subsection{Results}

The classification performance of the proposed HRAN and the three baseline models are shown as in Table~\ref{tab:results}. Having the attention mechanism appears to be effective and useful. With two-level attention mechanisms, the classification accuracy is greatly improved from $51.8\%$ (no attention) to $72.1\%$. Turn-level attention is found to be more effective than sub-turn-level attention, but the models with only one level of attention perform much worse than two-level attention. On the other hand, the results suggest that the speech from both the therapist and the client in the interaction contribute to predicting the observer-rated
empathy level.
% that is subjectively rated by the third-party observer.
% the results suggest that the observer's rating is determined by the speech of both the therapist and the client in the interaction.

It should be noted that the classification results on different test sets in the 6-fold CV have large variation. This may be due to very limited amount of data in our experiments. As a matter of fact, there are no more than 20 sessions in both development and test sets in each fold.

\begin{table}[htb]
  \caption{Classification accuracy on sessions with high vs. low level of therapist empathy}
  \label{tab:results}
  \centering
  \begin{tabular}{|c|c|c|}
  \hline
  Training speech                            & Network & Accuracy         \\ \hline
  \multirow{4}{*}{Both therapist and client} & HRN     & 51.8\%           \\ \cline{2-3} 
                                             & HRSAN   & 55.3\%           \\ \cline{2-3} 
                                             & HRTAN   & 64.4\%           \\ \cline{2-3} 
                                             & HRAN    & \textbf{72.1\%}  \\ \hline
  Therapist only                             & HRAN    & 61.9\%           \\ \hline
  Client only                                & HRAN    & 59.2\%            \\ \hline
  \end{tabular}
\end{table}

\section{Discussion}
\label{sec:disc}

\begin{figure}[htb]
  \centering
  \includegraphics[width=\linewidth]{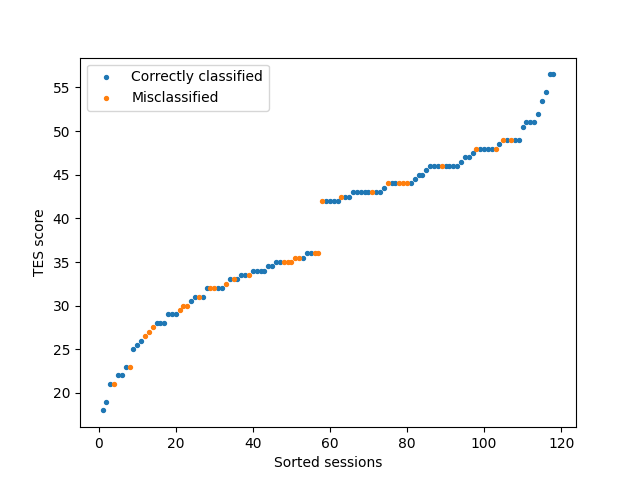}
  \caption{TES scores of the 118 counseling sessions in ascending order.}
  \label{fig:TES_score}
\end{figure}

Table~\ref{tab:con_mat} gives the confusion matrix on the 118 sessions involved in the CV experiments. There are 85 correctly classified sessions and 33 misclassified sessions. The number of low-empathy sessions misclassified as high-empathy is nearly two times of the misclassified high-empathy sessions. Figure~\ref{fig:TES_score} plots the TES scores in ascending order. It is noted that most of the misclassified low-empathy sessions have the scores over $30$, approaching the upper bound of the low-empathy group. Likewise, more than half of the misclassified high-empathy sessions have the scores below $44$, which is very close to the lower bound of the high-empathy group. These sessions could be viewed as with ``moderate" empathy level.

\begin{table}[htb]
  \caption{Confusion matrix of classification results}
  \label{tab:con_mat}
  \centering
  \begin{tabular}{|c|c|c|}
     \hline
     \diagbox{Ground truth}{Prediction} & High & Low \\ \hline
              High                      & 49   & 12   \\ \hline
              Low                       & 21   & 36  \\ \hline
  \end{tabular}
\end{table}

\begin{figure}[htb]
  \centering
  \includegraphics[width=\linewidth]{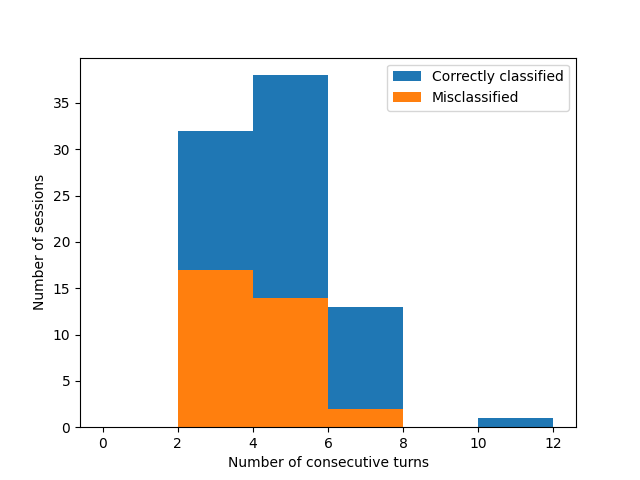}
  \caption{Statistics on the number of consecutive important turns in the 118 counseling sessions.}
  \label{fig:consecutive_turn}
\end{figure}

The attention weights assigned to individual speaker turns are analyzed to help us better understand how therapist empathy is realized and assessed in the counseling sessions. Speaker turns with attention weight values over a threshold are defined as ``important turns". This threshold is set to be the 70th percentile of the attention weights of all turns in a session. The important turns in each of the 118 sessions are extracted. Knowing that speaker interaction is an important aspect of conversation, the neighboring turns of each speaker turn are also considered. The number of important turns in neighborhood is counted. The average number of consecutive important turns in the session is computed. As shown in Figure~\ref{fig:consecutive_turn}, there are often 2 to 6 consecutive turns classified important turns in a session. This suggests that empathy assessment or detection is not done on isolated or disconnected speaker turns, but involves a few consecutive turns of interaction in the conversation.

\begin{figure}[htb]
  \centering
  \includegraphics[width=\linewidth]{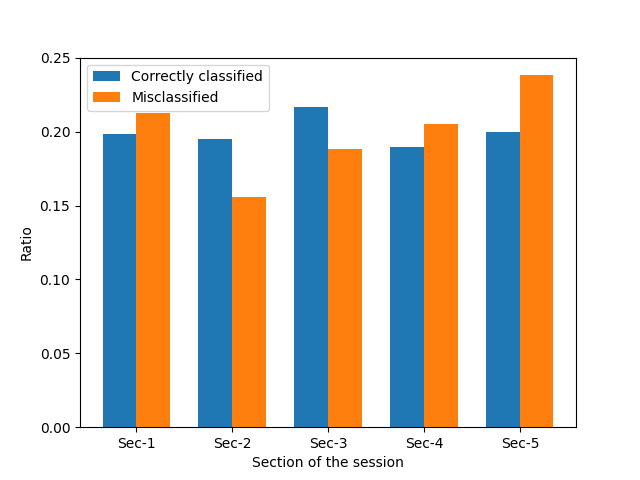}
  \caption{Average distribution of important turns in the five sections of correctly classified and misclassified sessions.}
  \label{fig:section_distribution}
\end{figure}

On the other hand, the distribution of important turns over the time course of a session is investigated. The session is divided into five sections that contain equal number of speaker turns. The percentage ratio of important turns in each section is computed. For each section, the average ratio across 118 sessions is computed. The percentage ratios of important turns in the five sections of counseling session are plotted as in Figure~\ref{fig:section_distribution}. It is noted that important turns tend to be more evenly distributed in the correctly classified sessions than in the mis-classified sessions. This suggests that the observer might have taken a holistic approach to determine the therapist's empathy level, rather than focus on a few specific parts of the conversation.

\section{Conclusions}
\label{sec:conc}

In this work, we propose to use the HRAN to determine the therapist's empathy level from conversational speech in counseling session. Benefiting from the two-level attention mechanisms, the model can achieve an accuracy of $72.1\%$ in classifying a counseling session as ``high" or ``low" empathy level, solely based on acoustic speech features. The results show that the speech from both the therapist and the client contribute to the observer's rating. By analyzing important turns with high attention weights, it is found that consecutive speaker turns jointly contribute to determining the empathy level. Analyzing the distribution of important turns over a session suggests that subjective empathy rating tends to take the whole session into consideration.

\section{Acknowledgements}
\label{sec:ack}
This research is partially supported by the Sustainable Research Fund of the Chinese University of Hong Kong (CUHK) and an ECS grant from the Hong Kong Research Grants Council (Ref.: 24604317).

\bibliographystyle{IEEEtran}

\bibliography{mybib}

% Generated by IEEEtran.bst, version: 1.13 (2008/09/30)
\begin{thebibliography}{10}
\providecommand{\url}[1]{#1}
\csname url@samestyle\endcsname
\providecommand{\newblock}{\relax}
\providecommand{\bibinfo}[2]{#2}
\providecommand{\BIBentrySTDinterwordspacing}{\spaceskip=0pt\relax}
\providecommand{\BIBentryALTinterwordstretchfactor}{4}
\providecommand{\BIBentryALTinterwordspacing}{\spaceskip=\fontdimen2\font plus
\BIBentryALTinterwordstretchfactor\fontdimen3\font minus
  \fontdimen4\font\relax}
\providecommand{\BIBforeignlanguage}[2]{{%
\expandafter\ifx\csname l@#1\endcsname\relax
\typeout{** WARNING: IEEEtran.bst: No hyphenation pattern has been}%
\typeout{** loaded for the language `#1'. Using the pattern for}%
\typeout{** the default language instead.}%
\else
\language=\csname l@#1\endcsname
\fi
#2}}
\providecommand{\BIBdecl}{\relax}
\BIBdecl

\bibitem{rogers1995way}
C.~R. Rogers, \emph{A way of being}.\hskip 1em plus 0.5em minus 0.4em\relax
  Houghton Mifflin Harcourt, 1995.

\bibitem{miller2009toward}
W.~R. Miller and G.~S. Rose, ``Toward a theory of motivational interviewing,''
  \emph{American Psychologist}, vol.~64, no.~6, p. 527, 2009.

\bibitem{elliott2011empathy}
R.~Elliott, A.~C. Bohart, J.~C. Watson, and L.~S. Greenberg, ``Empathy,''
  \emph{Psychotherapy}, vol.~48, no.~1, p.~43, 2011.

\bibitem{moyers2013low}
T.~B. Moyers and W.~R. Miller, ``Is low therapist empathy toxic?''
  \emph{Psychology of Addictive Behaviors}, vol.~27, no.~3, p. 878, 2013.

\bibitem{elliott2018therapist}
R.~Elliott, A.~C. Bohart, J.~C. Watson, and D.~Murphy, ``Therapist empathy and
  client outcome: An updated meta-analysis,'' \emph{Psychotherapy}, vol.~55,
  no.~4, p. 399, 2018.

\bibitem{xiao2012analyzing}
B.~Xiao, D.~Can, P.~G. Georgiou, D.~Atkins, and S.~S. Narayanan, ``Analyzing
  the language of therapist empathy in motivational interview based
  psychotherapy,'' in \emph{APSIPA ASC}, Dec. 2012, pp. 1--4.

\bibitem{chakravarthula2015assessing}
S.~N. Chakravarthula, B.~Xiao, Z.~E. Imel, D.~C. Atkins, and P.~G. Georgiou,
  ``Assessing empathy using static and dynamic behavior models based on
  therapist's language in addiction counseling,'' in \emph{Proc. INTERSPEECH},
  2015.

\bibitem{gibson2015predicting}
J.~Gibson, N.~Malandrakis, F.~Romero, D.~C. Atkins, and S.~S. Narayanan,
  ``Predicting therapist empathy in motivational interviews using language
  features inspired by psycholinguistic norms,'' in \emph{Proc. INTERSPEECH},
  2015.

\bibitem{gibson2016deep}
J.~Gibson, D.~Can, B.~Xiao, Z.~E. Imel, D.~C. Atkins, P.~Georgiou, and
  S.~Narayanan, ``A deep learning approach to modeling empathy in addiction
  counseling,'' \emph{Commitment}, vol. 111, p.~21, 2016.

\bibitem{xiao2013modeling}
B.~Xiao, P.~G. Georgiou, Z.~E. Imel, D.~C. Atkins, and S.~S. Narayanan,
  ``Modeling therapist empathy and vocal entrainment in drug addiction
  counseling,'' in \emph{Proc. INTERSPEECH}, 2013.

\bibitem{xiao2014modeling}
B.~Xiao, D.~Bone, M.~V. Segbroeck, Z.~E. Imel, D.~C. Atkins, P.~G. Georgiou,
  and S.~S. Narayanan, ``Modeling therapist empathy through prosody in drug
  addiction counseling,'' in \emph{Proc. INTERSPEECH}, 2014.

\bibitem{imel2014association}
Z.~E. Imel, J.~S. Barco, H.~J. Brown, B.~R. Baucom, J.~S. Baer, J.~C. Kircher,
  and D.~C. Atkins, ``The association of therapist empathy and synchrony in
  vocally encoded arousal,'' \emph{Journal of Counseling Psychology}, vol.~61,
  no.~1, p. 146, 2014.

\bibitem{xiao2015analyzing}
B.~Xiao, Z.~E. Imel, D.~C. Atkins, P.~G. Georgiou, and S.~S. Narayanan,
  ``Analyzing speech rate entrainment and its relation to therapist empathy in
  drug addiction counseling,'' in \emph{Proc. INTERSPEECH}, 2015.

\bibitem{bahdanau2014neural}
D.~Bahdanau, K.~Cho, and Y.~Bengio, ``Neural machine translation by jointly
  learning to align and translate,'' \emph{arXiv preprint arXiv:1409.0473},
  2014.

\bibitem{luong2015effective}
M.-T. Luong, H.~Pham, and C.~D. Manning, ``Effective approaches to
  attention-based neural machine translation,'' \emph{arXiv preprint
  arXiv:1508.04025}, 2015.

\bibitem{decker2014development}
S.~E. Decker, C.~Nich, K.~M. Carroll, and S.~Martino, ``Development of the
  therapist empathy scale,'' \emph{Behavioural and Cognitive Psychotherapy},
  vol.~42, no.~3, pp. 339--354, 2014.

\bibitem{cicchetti1994guidelines}
D.~V. Cicchetti, ``Guidelines, criteria, and rules of thumb for evaluating
  normed and standardized assessment instruments in psychology,''
  \emph{Psychological Assessment}, vol.~6, no.~4, pp. 284--290, 1994.

\bibitem{Tao2022CharacterizingTS}
D.~Tao, T.~Lee, H.~Chui, and S.~Luk, ``Characterizing therapist's speaking
  style in relation to empathy in psychotherapy,'' \emph{arXiv:2203.13127,
  submitted to INTERSPEECH}, 2022.

\bibitem{eyben2015geneva}
F.~Eyben, K.~R. Scherer, B.~W. Schuller, J.~Sundberg, E.~Andr{\'e}, C.~Busso,
  L.~Y. Devillers, J.~Epps, P.~Laukka, S.~S. Narayanan \emph{et~al.}, ``The
  geneva minimalistic acoustic parameter set (gemaps) for voice research and
  affective computing,'' \emph{IEEE Transactions on Affective Computing},
  vol.~7, no.~2, pp. 190--202, 2015.

\bibitem{Xue2019}
\BIBentryALTinterwordspacing
W.~Xue, C.~Cucchiarini, R.~van Hout, and H.~Strik, ``Acoustic correlates of
  speech intelligibility: the usability of the egemaps feature set for atypical
  speech,'' in \emph{Proc. SLaTE 2019: 8th ISCA Workshop on Speech and Language
  Technology in Education}, 2019, pp. 48--52. [Online]. Available:
  \url{http://dx.doi.org/10.21437/SLaTE.2019-9}
\BIBentrySTDinterwordspacing

\bibitem{haider2019assessment}
F.~Haider, S.~De~La~Fuente, and S.~Luz, ``An assessment of paralinguistic
  acoustic features for detection of alzheimer's dementia in spontaneous
  speech,'' \emph{IEEE Journal of Selected Topics in Signal Processing},
  vol.~14, no.~2, pp. 272--281, 2019.

\bibitem{eyben2010opensmile}
F.~Eyben, M.~W{\"o}llmer, and B.~Schuller, ``Opensmile: the munich versatile
  and fast open-source audio feature extractor,'' in \emph{Proc. ACM Multimedia
  (MM)}, 2010, pp. 1459--1462.

\bibitem{cho2014properties}
K.~Cho, B.~Van~Merri{\"e}nboer, D.~Bahdanau, and Y.~Bengio, ``On the properties
  of neural machine translation: Encoder-decoder approaches,'' \emph{arXiv
  preprint arXiv:1409.1259}, 2014.

\bibitem{sukhbaatar2015end}
S.~Sukhbaatar, J.~Weston, R.~Fergus \emph{et~al.}, ``End-to-end memory
  networks,'' \emph{Advances in Neural Information Processing Systems},
  vol.~28, 2015.

\end{thebibliography}

\end{document}